# The wavelength of light as Thomas Young invented it

Olivier Morizot
*Aix-Marseille Univ, CNRS, Centre Gilles Gaston Granger, Aix-en-Provence, France.*
Olivier.morizot@univ-amu.fr
ORCID ID: 0000-0002-5640-8430

This preprint has not undergone peer review or any post-submission improvements or corrections. The Version of Record of this article is published in *Archive for History of Exact Sciences* 80 (2026) and is available online at https://doi.org/10.1007/s00407-026-00387-7

**Abstract**
This paper documents Thomas Young's invention of a concept which he alternately refers to as the "magnitude", "breadth", "interval" or "length of an undulation" of light. First, the paper highlights that Young was first to link some concept of a wavelength to a theory of optics and to assign a precise value to this length for each component of the colour spectrum. Then, it sets out an explanation of how Young made up these values. Finally, it examines the reasons why Young introduced an optical wavelength into his theory of light, insofar as these reasons might in turn shed light on why he was first to do so.

**Keywords**: Thomas Young; history of optics; concept; wavelength; interference; light.

## 1. Introduction

This article aims first and foremost to document, as a historical fact, Thomas Young's invention of a concept he alternately termed the "magnitude", the "breadth", the "interval" or the "length of an undulation" of light. I shall thus point out that Young was first not only to link such a concept to a theory of optics, but also to assign precise values – actually comparable to those currently accepted – to the physical lengths characterizing the different components of the coloured spectrum.

In doing so, I shall not claim that Young's invention is identical to the current concept of optical wavelength, let alone it is the *point of origin* from which the current concept was forged, at least since Young's concept of an optical "undulation" – as we shall see – is significantly different from the actual one, and since there is always something artificial about assigning a single and definite origin to a concept, or to a theory. Instead, I shall simply set out the concept of "length of an undulation" of light as it was introduced by Young, endeavouring to convey its precise meaning and the specific role this concept plays in his optical theory. I shall not, therefore, be presenting anything historically original at first, since the information I shall start to recall is explicitly set out in the text of his "Theory of Light and Colours", published in 1802. However, this exposition will not be worthless as, by focusing primarily on Young's invention of the law of light interference (Cantor 1983; Kipnis 1991; Darrigol 2009), historiography seems to have too rarely emphasised his concurrent invention of the length of undulation of light until now.

This text will then attempt to explain how Young obtained his initial values for the lengths of monochromatic undulations. This point too is apparently not a matter of revealing a historical secret, since Young publicly announces – albeit briefly – the method he employed in the very text of his "Theory of Light and Colours". My aim, however, is to go a bit further and ascertain to what extent what he claims he has done actually enabled him – or not – to arrive at the results he ultimately presented.

I shall then conclude by considering, in a more speculative vein, the reasons why Young introduced the length of optical undulation, insofar as these reasons might in turn shed light on why he was first to do so. I shall therefore argue that Young invented the concept of "length of an undulation" of light because he was the first to need it – in this instance, to put to the test his brand-new law of interference of light. To this end, I shall endeavour, throughout the text, to place the optical problems Young was seeking to resolve in perspective alongside those formulated by Descartes, Newton, Hooke, Huygens, Malebranche and Euler before him.

The conclusions presented here are therefore the deliberate outcome of an essentially internalist and conceptual approach to the history of science, in that I shall focus on scientific texts, on their internal structure, on the meaning and scope locally attributed to certain concepts within these structures, and on the contrasts emerging between these concepts and those set out in other texts and by other authors.

## 2. Undulations of light

Between January 1800 and November 1803, Young presented five papers to the Royal Society of London relating to his theory of light (Young 1800; 1801; 1802a; 1802b; 1804). Strictly speaking, it would therefore be appropriate to analyse these five papers jointly in order to gain the best understanding of the meaning of Young's optical theory at that time. However, I shall take the liberty of focusing here on the central paper in this series, entitled "On the Theory of Light and Colours", in which Young specifically applies the term "length of an undulation" to light and, for the first time in history, assigns a value of this length to each component of the spectrum.

Unusually for the time, Young's "Theory" is introduced by four hypotheses, the implications of which are then set out to test. These hypotheses are: I. That "A luminiferous Ether pervades the Universe, rare and elastic in a high degree"; II. That "*Undulations* are excited in this Ether whenever a Body becomes luminous"; III. That "The Sensation of different Colours depends on the frequency of *Vibrations*, excited by Light in the Retina"; and IV. That "All material Bodies have an Attraction for the ethereal Medium, by means of which it is accumulated within their Substance, and for a small Distance around them, in a State of greater Density but not of greater Elasticity" (Young 1802a: 14–22)[1].

From this, one must above all bear in mind that, according to Young, colours are first and foremost sensations, resulting from "vibrations" of the retina communicated to the *sensorium* along the optical filaments, which are themselves caused by longitudinal "undulations" of a subtle and elastic medium occupying all space – including the interior of material bodies – called "ether", and that these "undulations" of the ether are what Young calls "Light".

Note that the term "undulation" is most important here, as, according to Young, the term explicitly refers to a movement that is "continued alternately" and which ceases immediately when the excitation that caused it ceases. It then differs from a "vibration", that is also a continuous alternating movement, but which will go on over some time once its excitation has ceased. As a set of examples, Young states that the "vibrations" of a musical string, which continue long after it has been struck, produce "undulations" in the air that cease as soon as the string's "vibration" ceases; and that "undulations" of the ether are produced by the "vibrations" of the corpuscles of luminous bodies – and therefore shall stop as soon as those bodies stop shining – and, in turn, produce "vibrations" in the retina, as evidenced by the brief persistence of vision after its excitation has ceased.

[1] Words "undulations" and "vibrations" were stressed in italics by the author of the present paper.

Young, however, never speaks categorically about the temporal form of the undulation of ether: at most, in his first article on sound and light he draws and combines sawtooth-shaped undulations, but he states that he chose this form for the sake of simplicity and that these undulations might just as well be considered to present a "harmonic" profile (Young 1800); and that the latter solution would even facilitate certain calculations. He also devised a wooden instrument for visualising the superposition of two sinusoidal undulations and illustrate some consequences of his principle of interference (Young 1802c) but, at least in the texts published between 1800 and 1803, in which he progressively develops his optical theory pretty much on his own, Young never takes a definitive stance on the form of the undulation of light, probably out of a concern not to mistake a particular case for the general one. Nevertheless, it is clear to him that light is the product of a continuous and periodic undulation of the ether – whatever the temporal profile of this undulation may be.

I emphasise this point because behind the term "undulation" already lies one of the preliminary conditions enabling the invention of the concept of a length of undulation. Indeed, not only could this concept certainly not have emerged within the framework of a theory of luminiferous projectiles, such as those that proliferated in the 18th century in the wake of Newtonian optics (although we shall soon qualify this statement), but nor could it have arisen within Descartes' medium theory, in which light is the consequence of a pressure exerted statically and radially from the surface of luminous bodies and instantly transmitted through an absolutely compact ether (Descartes 1637); nor within Hooke's theory, which took no interest in the possibility of a periodicity of the ether's vibrations (Hooke 1665); nor within Huygens's theory, which envisaged light as the result of random percussions produced by the corpuscles of luminous bodies and thus propagating through the ether with no relation of periodicity (Huygens 1690); nor in Malebranche's work, who, although he was the first to assign a certain temporal "promptitude" of ether vibration to each component of the colour spectrum, was nevertheless unable to conceptualise the spatial period of this vibration as he posited that light propagated instantaneously (Malebranche 1699; 1712); nor, finally, within Euler's, who envisaged light as a series of pulses propagating independently at the same finite speed, and periodically separated by a time interval during which the ether was at rest, the value of which interval varied with colour (Euler 1746)[2].

Focusing on Euler's case only, then – for he certainly produced the most advanced theory of the optical medium prior to the publication of Young's – he could not introduce any concept of wavelength, at least because light was not regarded as an undulation in his theory. This interval of rest separating two successive pulses in Euler's work – he called it "intervallem", in Latin – can obviously not be a wavelength, since there is no undulatory motion in this interval. According to Euler, the particules of ether undergo an undulating movement indeed, but this undulation wholly lies within the very thickness of each pulse, each consisting of a single period of undulation of the ether itself extending over a length $2\alpha$ – referred to as "magnitudo" in Latin – much shorter that the interval separating two pulses and which is the same for all components of the spectrum. Incidentally, Euler estimates that the length of an "intervallem", varying with colour, to be greater than 600 ft. (Paris) – that is approximately 190 m – and to decrease from violet to red[3].

Thus, there is no trace in optics of any semblance of a wavelength prior to Young's work. Except, perhaps, in the writings of an author whose authority was beyond doubt, but who expressly rejected the idea that light could be propagated as a wave. Indeed, whilst Newton attributed the sensation of light and colour to the reception on the retina of material projectiles emitted from luminous bodies, he nevertheless suggested, in the concluding queries of his *Opticks*, that the various sensations of colour one can experience were the result of vibrations in the ether that is contained within the optic nerve, which were caused by the impact of these projectiles. The diversity of sensations of pure colours would then reflect the diversity of "bignesses" of these vibrations, depending in turn on the diversity of bulk of these projectiles impinging the retina (Newton 1730: 328). As for the exact meaning Newton attributed to the term "bigness" – which is probably not that of a spatial period of an undulation – much has already been said (Sabra 1963; Blay 1980), but this original meaning is less important in our case than the way in which Young himself could have interpreted it. Similarly, Young would certainly have been struck by this letter to the Royal Society in which Newton suggested to Hooke that the various components whose presence he had revealed in the spectrum of white light might be compatible with Hooke's vibrational conception of light, provided one considered, for example, that light itself consisted of vibrations of the ether of different "bignesses" already; vibrations which Newton, moreover, estimated to repeat at "less distance than the hundred thousandth part of an inch" (Newton 1672: 251), or approximately 250 nm.

[2] For a more detailed comparison of these various kinetic models of light vibration, see (Shapiro 1973).
[3] For a more detailed discussion of Euler's *Nova theoria* see (Hakfoort 1995).

But in Newton's case, although a concept approaching that of wavelength seems on the verge of emerging, the potential of this concept is not realised, as its author simply does not himself conceive of the possibility of light as a wave. Nevertheless, it is undeniable that these texts by Newton inspired Young's invention (Young 1802: 17–20; 40) and that they were the first to assign to the characteristic spatial period of the undulation of light an order of magnitude comparable to that attributed to it today.

To summarise, whilst the concept of wavelength was absent from optics prior to Young, it is also worth noting that the very idea of a periodic and continuous light undulation was rarely proposed either; which in turn explains the absence of a concept that might have described its spatial period.

### 3. Breadth, magnitude, length and interval

Let us then return to Young's text and observe how he introduces this concept of the "length of an undulation". In fact, the theory he proposes – which is based on the four hypotheses already mentioned – derives most of its value, he says, from the fact that it is able to reduce a considerable number of phenomena of optics (and of nature, actually), hitherto regarded as disparate, to a single explanation (Young 1802a: 12). This single explanation relates, firstly – as has been noted – to the idea that light is the result of an undulation of the ether. Secondly, according to the eighth proposition of his text, this explanation relates to the fact that: "When two Undulations, from different Origins, coincide either perfectly or very nearly in Direction, their joint effect is a combination of the Motions belonging to each" (Young 1802a: 34). Or, in words Young would use a few months later in order to clarify his point, that: "wherever two portions of the same light arrive at the eye by different routes, either exactly or very nearly in the same direction, the light becomes most intense when the difference of the routes is any multiple of a certain length, and least intense in the intermediate state of the interfering portions; and this length is different for light of different colours" (Young 1802d: 387).

This is not the place to discuss the reasons why this second formulation of the law of interference of light no longer contains the word "undulation" – this has already been done elsewhere (Kipnis 1991) – yet it will not have escaped one's attention that this second formulation contains the word "length", which is found as the root of the modern term "wavelength", as well as in the more archaic expression "length of undulation". Note, moreover, that when considering all five articles in which Young progressively sets out his optical theory (Young 1800; 1801; 1802a; 1802b; 1804), the word "length" is used only five times in order to describe the spatial period of the undulation of ether. In 20% of cases (that is, on seven occasions), Young prefers the term "breadth" (Young 1802a: 26; 35; 38; 42); in 15% of cases (on five occasions) he writes "magnitude" instead (Young 1802a: 21, 38; 1802b: 393) and in 50% of cases (on 17 occasions) he uses the term "interval" – but these latter instances all occur in the last of these five texts, where Young specifically attempts to conceal the wave-like nature he attributes to light (Young 1804: 4, 5, 6, 7, 12). Thus, in only 15% of cases (i.e. on 5 occasions) does Young write "length" in order to address the spatial period of the undulation (Young 1800: 129; 1802a: 39; 1802b: 387). And I believe this sheds light on the fact that Young is indeed in the process of developing a new concept for which there is no pre-established terminology designing the physical entity it is trying to point at.

We can therefore see Young drawing alternately on the terminology of Euler, who – as was mentioned earlier – spoke of ether pulses of a given "magnitudo", separated by variable "intervalla" (Euler 1746); but he also draws on Newton, who, in the *Principia*, on the one hand, defined the "*breadth* of the waves [..] the transverse measure lying between the deepest part of the hollows, or the tops of the ridges" (Newton 1729, II: 172) in the course of his study of waves propagating across the surface of water; and who, on the other hand, devoted several pages to the study of "Pulses of the air, by which sounds are propagated, their *intervals* or *breadths* determined […]; these intervals in sounds made by open pipes probably equal to twice the *length* of the pipes" (Newton 1729, II: Index)[4], that is a model of sound that Euler would actually later adopt in his acoustic theory, before transposing it to his theory of optics.

The variability in the terminology used by Young in order to describe the spatial repetition period of his undulations of light thus highlights not only the absence in optics of a term that met his requirements prior to his work, but also his effort to draw upon other fields than optics – in this case, acoustics and hydrodynamics – and import concepts from them that could both resemble the new optical concept he was attempting to forge and satisfy the understanding of his contemporaries, who, faced with the novelty of his optics, might rely on their prior knowledge of hydrodynamic, acoustic or optical theories – notably those of Newton and Euler – in order to grasp its meaning.

By the way, the close articulation of optics to acoustics and hydrodynamics is certainly not specific to Young's work (Darrigol 2009; 2010; 2012). However, it is all the more natural to him since his whole optical theory is based on a close and material analogy between light and sound (Young 1800), whose theories are in turn both supported by the same hydrodynamical principles (Young 1807).

### 4. The length of an undulation, in air

But let us focus on the five instances of the term "length", taken to mean the spatial period of an optical undulation, in these five early optical papers of Young. Two of these instances – the last two, in fact – appear in the sentence summarising his law of interference, dated July 1802, which we have already quoted (Young 1802b: 387). Meanwhile, the first occurrence of the term is found in the first paper of this series, in a passage where Young explicitly examines the analogy between light and sound, and then states: “Now this is precisely similar to the production of the same sound, by means of an uniform blast, from organ-pipes which are different multiples of the same length. […]. It would seem, that the determination of a portion of the track of a ray of light through any homogeneous stratum of ether, is sufficient to establish a length as a basis for colorific vibrations” (Young 1800: 129). Here though, I would only take into account the second occurrence of term “length”, since the first refers not to light but to an organ pipe. Yet precisely, this passage increases then the likelihood of the incorporation into optics of the term “length”, borrowed from the acoustic theory of sounds produced by organ pipes; that Newton had already shown – as we have just seen – to be characterized by a spatial period – which he called an “interval” – that was an integer multiple of the “length” of these pipes (Newton 1729, II: Index).

[4] Words "breadth", "interval" and "length" were stressed in italics by the author of the present paper.

Then, let us finally turn now to the only two instances in which Young explicitly refers to the “length of an undulation” that can be found in his “Theory of Light and Colours”, where he states that “The absolute length and frequency of each vibration [of light] is expressed in the table; supposing light to travel in $8_{1/8}$ minutes 500,000 000 000 feet” (Young 1802a: 39). The table in question (Figure 1) does indeed expose, in its second column, the “Length of an undulation, in parts of an inch, in Air”, which turns out to be the one and only occurrence of the fully expanded expression “length of an undulation” in these early articles of Young’s. This second column therefore assigns to each principal colour of the spectrum (as they were defined by Newton) listed in the first column, as well as to each of the boundaries marking the transition between two of these colours in the spectrum – which Young qualifies as “intermediate” colours – a value, in “parts of an inch”, of the spatial period of the undulation of ether causing the corresponding sensation. Hereafter, I shall denote this “length” by the letter $d$ rather than $\lambda$, so as to avoid projecting onto it the knowledge subsequently developed regarding the optical wavelength.

| *Length of an Undulation in nm, in Air* | Colours. | Length of an Undulation in parts of an Inch, in Air. | Number of Undulations in an Inch. | Number of Undulations in a Second. |
|---|---|---|---|---|
| 675 | Extreme - | .0000266 | 37640 | 463 millions of millions |
| 648 | Red - - | .0000256 | 39180 | 482 |
| 624 | Intermediate | .0000246 | 40720 | 501 |
| 610 | Orange - - | .0000240 | 41610 | 512 |
| 598 | Intermediate | .0000235 | 42510 | 523 |
| 577 | Yellow - | .0000227 | 44000 | 542 |
| 557 | Intermediate | .0000219 | 45600 | 561 (= $2^{48}$ nearly) |
| 535 | Green - - | .0000211 | 47460 | 584 |
| 515 | Intermediate | .0000203 | 49320 | 607 |
| 497 | Blue - - | .0000196 | 51110 | 629 |
| 480 | Intermediate | .0000189 | 52910 | 652 |
| 470 | Indigo - - | .0000185 | 54070 | 665 |
| 460 | Intermediate | .0000181 | 55240 | 680 |
| 442 | Violet - - | .0000174 | 57490 | 707 |
| 425 | Extreme - - | .0000167 | 59750 | 735 |

**Figure 1** – Each of the seven principal components of the spectrum and the boundaries between these colours, as identified by Newton (column 1), is associated with a length (column 2), a wave number (column 3) and a temporal frequency (column 4) (Young 1802a: 39). To the left of the table, I added a column exhibiting the value in nm of “the length of an undulation” computed by Young.

I have also converted the values of $d$ into nm in a column I have added to the left of the table, so that one can more easily compare these values provided by Young with those currently used for optical wavelengths. According to Young, they range from 676 nm for extreme red to 424 nm for extreme violet. The correspondence between these very first measurements of the lengths of the undulations of light in history and the presently accepted values ranging approximately from 700 nm for "extreme red" to 400 nm for "extreme violet" is all the more remarkable given that these physical measurements quantify a colour sensation that necessarily varies from one individual to another, so that those values are presently to be considered as rough averages over a wide range of individuals; although, as we shall see, Young's values actually quantify Newton's colour sensations only.

It should also be added that the third column of this table simply shows the number of undulations corresponding to each colour that could be counted within the space of one inch, that is to say, the reciprocal $1/d$ of each value given in column two – a parameter which in modern optics would be called the wavenumber. Then, the last column shows the temporal frequency of each undulation – which I shall denote by $f$ – calculated by dividing the speed of light $c$ (here taken as $8_{1/8}$ minutes to travel the distance from the Sun to the Earth, i.e. 313,000 $km \cdot s^{-1}$) by the length $d$ of the undulation: $f = c/d$.

Interestingly, Young admits that he did indeed obtain these first experimental values for optical lengths without carrying out a single experiment. He even invites his readers to place all the more confidence in his results precisely because they were deduced from experiments that were neither conceived, nor carried out, within the framework of his own theory. It is actually well known that in his *Opticks*, Newton extensively described and analysed the coloured rings that appear when one illuminates an optical system consisting of a convex lens placed on a flat glass surface. In particular, Newton noted that these rings, which appeared iridescent when the system was illuminated with white light, transformed into a periodic series of monochromatic rings when the system was illuminated by a single component of the spectrum. He then observed that the periodicity of these rings corresponded precisely to the arithmetic progression of the thickness of the air trapped between the lens and the glass: for transmitted light, for instance, the rings were dark when the thickness was $L$, $3L$, $5L$, etc., and bright for even multiples of $L$. Finally, Newton noted that the value of $L$ increased as the colour of the pure light illuminating the system varied from violet to the extreme red of the spectrum (Newton 1730).

Whatever explanation Newton gave for this phenomenon in terms of "fits of easy transmission" or "reflection" (Shapiro 1993), what ultimately matters most to us in this case is that his text provided the precise value of this thickness $L$ for all the coloured components of the spectrum. Or, to be more precise, Newton gave the exact ratios in which these thicknesses were to be found: "the thicknesses of the Air between the Glasses there, where the Rings are successively made by the limits of the seven Colours, red, orange, yellow, green, blue, indigo, violet in order, are to one another as the Cube Roots of the Squares of the eight lengths of a Chord, which found the Notes in an eighth, *sol*, *la*, *fa*, *sol*, *la*, *mi*, *fa*, *sol* ; that is, as the Cube Roots of, the Squares of the Numbers, 1, 8/9, ¾, 5/6, 2/3, 3/5, 9/16, ½" (Newton 1730: 186). That is to say, if the thickness $L$ corresponding to the first dark ring in extreme red light were 10,000, the thicknesses of that same first ring for a system illuminated by a spectral component situated exactly at the boundary between two principal colours – i.e. by an "intermediate" component, so as to use Young's term – would be, in turn, "9243 [for the red-orange boundary], 8855 [for the orange–yellow boundary, then] 8255, 7631, 7114, 6814 and 6300 [for the subsequent ones]" (Newton 1730: 200).

One hundred years later, Young offered a new interpretation of the phenomenon of the coloured rings observed by Newton, using the law of interference of light he had just invented (Young 1802a: 37–41): at every point on the glass surface where the undulation of light is likely to continue its path downwards, part of that undulation is reflected upwards (since any transparent surface always produces partial reflection), and then downwards again when it returns to the lower surface of the lens; thus, to the first undulation, which ought to be transmitted directly, eventually overlaps with a second undulation that has been reflected twice. Now, according to Young, if the additional distance travelled through the air by this doubly reflected undulation corresponds exactly to the "length" covering one period of the undulation, or to an integer multiple of that length, the two undulatory motions of the ether will add together and produce an increase in local brightness. But if this additional distance – simply equal to twice the local thickness $2L$ at normal incidence – is equal to an integer multiple of $d/2$ (half the "length" of a period of the longitudinal undulation), any forward motion of one will superimpose on a retrograde motion of the other, leading to a drop in brightness and thus to the local observation of a dark ring. Consequently, Young implies that, in order to establish the series of lengths $d$ of the undulations of ether, he simply needs to collect the thickness $L$ assigned by Newton to the first dark ring in transmission for each spectral component, and to divide this thickness by 4: at this point, the round-trip length $2L$ should indeed be equal to half the length of the undulation, $d/2$.

## 5. Assigning a length to each monochromatic undulation

I have said – because he suggests it himself – that Young merely applied the preceding line of reasoning to Newton's experimental data[5] . However, the manuscript of his "Theory of Light and Colours" (Young 1802d) held in the archives of the Royal Society (Figure 2), indicates a more complex process, to which Young was at least partly condemned by Newton's text.

| COLOURS | Length of an undulation in parts of an inch, in air. | Number of undulations in an inch | Number of undulations in a second. |
|---|---|---|---|
| Extreme | .0000266 | 37640 | 463 millions of millions |
| Red | .0000256 | 39180 | 482 |
| Intermediate | .0000246 | 40720 | 501 |
| Orange | .0000240 | 41610 | 512 |
| Intermediate | .0000235 | 42510 | 523 |
| Yellow | .0000227 | 44000 | 542 |
| Intermediate | .0000219 | 45600 | 561 |
| Green | .0000211 | 47460 | 584 |
| Intermediate | .0000203 | 49320 | 607 |
| Blue | .0000196 | 51110 | 629 |
| Intermediate | .0000189 | 52910 | 652 |
| Indigo | .0000185 | 54070 | 665 |
| Intermediate | .0000181 | 55240 | 680 |
| Violet | .0000174 | 57490 | 707 |
| Extreme | .0000167 | 59750 | 735 |

**Figure 2** – Table of lengths (column 2), number of undulations in an inch (column 3) and temporal frequencies (column 4) associated with the coloured components of the spectrum (column 1) as it appears in the manuscript version of "The Theory of

[5] To be precise, Young states that he deduced these lengths "from Newton's measures of the thicknesses reflecting the different colours" (Young 1802a: 38). The reasoning and the series of ratios relating the thicknesses $L$ remain, however, exactly the same in the case of reflected light as in that of the rings in trasmitted light, which I have chosen to discuss here; the case of reflection, however, introduces a shift of half the length of the undulation at the air/glass interface, as Young notes, which allows the reflection pattern to be regarded as complementary to that in transmission. Still, for sake of simplicity I have not deemed it necessary to comment on this shift here and this is mainly why I chose to develop the whole demonstration on the equivalent case of transmitted light.

Light and Colour" (Young 1802d: f.29). To be compared with Figure 1. The present paper will focus however mainly focus on the series of numerical values which were crossed out in the middle of the document.

Indeed, the passage from Newton's *Opticks* that has been cited above (1730: 186) did not provide the exact thicknesses of air corresponding to the first dark ring in transmission for each colour, but only the ratio of these thicknesses to one another, and this applied solely to the "intermediate" spectral components (not to the principal colours themselves). Consequently, in order to determine the value of the thickness $L$ corresponding to the first dark ring in each of these intermediate colours, it was necessary to know at least one absolute thickness of air that could serve as a standard, so as to deduce the values of all the other thicknesses by means of these ratios. For that matter, in his *Opticks*, Newton provided a single thickness of air assigned to a bright ring observed in reflected monochromatic light, $L$ = 1/178,000 of an inch; which in turn corresponds to the thickness of air demonstrating the first dark ring in transmission for a "bright citrine yellow, or confine of yellow and orange" (Newton 1730: 234). The colour corresponding to the only thickness of air actually given by Newton was therefore designated in a surprisingly vague manner, although one might initially assume that, since it was for the intermediate colours that Newton gave a series of ratios of thicknesses, it was also to an intermediate colour – namely the "confine of yellow and orange" – that he assigned the absolute thickness which would serve as the standard for determining all the others.

Consequently, for this intermediate colour, one should certainly be able to read in the third of Young's column a wavenumber $1/d = 4/L$ = 44,500 inch$^{-1}$ and therefore, in the second column, an undulation length $d = L/4$ = 0.0000225 inch. However, neither of these two values appears in the handwritten table in Figure 2, which was faithfully reproduced in Young's publication (Figure 1).

Unless one takes a closer look at a series of values that was written in and then crossed out in the middle of Figure 2. In this central column, which Young had evidently labelled "UNDULATIONS IN AN INCH IN AIR" before crossing it out, and which should therefore correspond to a preliminary version of the third column of Figure 1, one does find indeed, opposite "Yellow" – which Young evidently chose to interpret as Newton's "citrine yellow" rather than the "confine of yellow and orange" – a value "44500" inch$^{-1}$, corresponding precisely to a thickness $L$ of 1/178,000 of an inch. Then, from this value – which, in this striped column, is actually the one exhibiting the fewest significant figures – Young seems to have deduced the wavenumbers of all the other principal components of the spectrum (red, orange, green, blue, indigo and violet) by multiplying it by the thickness ratios provided by Newton. Or, to be more precise, since Newton rather gave the ratios of the thicknesses corresponding to the boundaries separating these colours (1730: 200), Young did not use the original series, but a series of ratios he calculated himself, by averaging, in pairs, the successive values from Newton's list: as a matter of fact, if one calculates the ratio of successive values in the series of wave numbers that was striated, one obtains exactly the same series as would be obtained by averaging the successive values in Newton's series.

Therefore, Young did not derive his results directly from Newton's data. Rather, he interpreted them in at least three ways. First, he interpreted them conceptually, by converting thicknesses of air into multiples of the periodic length of an undulation thanks to both his concept of undulations of ether and his idea of their interference. Second, he interpreted these data qualitatively, by assigning Newton's single absolute thickness to "Yellow" rather than to the boundary between yellow and orange, as Newton suggested. Third, he interpreted them numerically, by recalculating an entire series of ratios on the assumption that each principal colour occupied exactly the midpoint between its two boundaries.

That being said, for a reason I have not yet discovered the situation is actually even more complex, since this initial result obtained by Young was eventually crossed out, and since he replaced it with a second set of wavenumbers, which he most likely obtained in the same way. That is, starting again from a value for the wavenumber corresponding to "Yellow", however taken it this time as equal to 44,000 inch$^{-1}$ (rather than 44,500) and extending it, first, to the other principal colours using his own series of ratios (calculated as the averages of Newton's successive ratios for the intermediate colours); second, to the "intermediate" colours. Indeed, in a second step, Young increased the number of values included in his list of wavenumbers $1/d$ by assigning one to each of the intermediate colours, starting from the value he had just obtained for extreme red through his own series of ratios, and then simply using Newton's series which was precisely made for that purpose. This is, in any case, what can be inferred from a series of clues, such as the very distinct colour of the ink that was used in order to inscribe all the words "intermediate" in the first column of the manuscript, as well as the slightly smaller size of the figures indicating the wavenumbers of these intermediate colours – as if each of their values had been inserted at a later stage between two pre-written values for the principal colours. Finally, Young would have converted these values of $4/L$ into those of $d = L/4$ and $f = 4c/L$, so as to fill in the second and fourth columns of the table and arrive at these values for lengths and frequencies in the manuscript, which are found today in the printed version of the "Theory of Light and Colours" (Figure 1).

## 6. Leads and Conclusions

I would certainly like to be able to explain why Young ultimately chose to reject the only experimental value for the thickness of air that had been provided by Newton, even though, at first, he had taken it into account. I have put forward several hypotheses in this regard, the testing of which has so far always proved unsatisfactory: I have indeed assumed that Young, although he often claimed to be reluctant to carry out his own experiments, had himself reproduced Newton's rings experiment and derived from it a value of 1/176,000 of an inch for the thickness of the first dark ring in transmission (or bright ring in reflexion) in yellow light – without mentioning it in his papers and without any evidence of this in his archives. But I have also calculated that, in order to distinguish that value new value from 1/178,000 of an inch, on a system comparable to that used by Newton, one would need to be able to measure the radius of these rings to an accuracy of one thousandth of an inch, whereas Newton himself stated that he could not make this measurement to better than one hundredth. So that, claiming to be capable of carrying out measurements ten times more precise than Newton's, and to regard them as so much more reliable as to use them without ever mentioning it – at the risk of people realising the substitution – does not seem consistent with what I know of Young.

I have also speculated that he might have determined the new value of his standard wavenumber using another optical experiment; such as that involving the "striated surfaces" which he actually carried out and analysed in his "Theory of Light and Colours" just before discussing the case of the rings (Young 1802a: 35–37). This experiment involves the diffraction of red light by a kind of diffraction grating, the pitch of which he knew, so that it could indeed have yielded – had Young chosen to interpret the data in this way – a wave number of 39,750 inch$^{-1}$ for red light. However, this value turns out to be closer to the value Young had initially recorded (which was 39,570 inch$^{-1}$) than to the one he ultimately published instead (39,180 inch$^{-1}$). It is therefore unreasonable to attribute the correction he made to the single value for air thickness provided by Newton solely to the inclusion of data from this experiment on striated surfaces.

The current incompleteness of Young's archives relating to the period prior to the publication of his "Theory of Light and Colours" therefore leaves me at a loss as to the reasons that may have led him to assign this particular value of 44,000 to the "Number of undulations [of ether] in an inch" causing the sensation of yellow, and consequently to assign the corresponding value of 0.0000227 inches (or 577 nm) to the "length of an undulation" of yellow light, from which he deduced all the others; to such an extent that I presently cannot rule out the possibility that the actual cause for this change would be an aesthetic or pragmatic choice, or even a mere oversight.

For the time being, I shall therefore conclude this paper, hoping only that I have succeeded in suggesting that if Thomas Young – and no one before him – invented and immediately quantified the length of the optical undulation, it is because he did not simply propose a concept that would make it possible to distinguish the components of the spectrum – just as Newton had pointed out to Hooke the "bigness" of the vibrations, as Malebranche had focused on their "promptitude", or as Euler had modelled the rate of repetition of pulses travelling periodically through the ether by means of his "intervalla". On the contrary, Young was first to achieve this dual realisation, both because he conceived light as an actual undulation, and because he required a length characterising this undulation that would enable him to apply his new law of interference to the explanation of coloured phenomena. Once the concept of a continuous and periodic undulation had been established, and once the ability of two undulations to interfere had been posited, Young needed indeed to establish the length of one period of the undulation as a standard against which to compare the differences in the optical paths that had been travelled by the two undulations interfering, so that he could finally determine their state of interference at the location of their crossing. In this respect, the invention of the length of an undulation of light can therefore be interpreted as a direct consequence of the invention of the law of interference of light, as is corroborated by the conjoint emergence of these two entangled concepts in Young's writings.

## 7. Postscript: Fresnel's optical wavelength

Ten years later, Fresnel reintroduced the concept of an optical length of an undulation into his own optical theory, the success and impact of which ensured the definitive integration of a concept of wavelength into optics where, despite the alterations it would continue to undergo, it still plays a central role today. Actually, this concept went through several transformations along with the progression of Fresnel's writings already, which, in retrospect, bear witness to the influence of Young's invention, despite the far lesser impact that his theory had on the Continent as well as in the British Isles, as compared to Fresnel's (Cantor 1983).

In Fresnel's "premier mémoire sur la diffraction", dated October 1815, written at a time when he did not yet have an explanatory model for the colours of thin plates and was not yet familiar with Young's work, Fresnel developed an initial geometric model of the diffraction of white light, in which he assigned a value of 516.7 nm to the average length of the undulations of a mechanic medium; a value which he justified through an elaborated average of the same data from Newton's *Opticks* that we have referred to in relation to Young (Fresnel 1866, I: 18; 98). I believe, however, that Fresnel forged this value backwards, as the one that allowed the best correspondence between the elegant algebraic formula he had devised to model the fringe spacing in monochromatic light – a formula which, of course, involves the length of a period of the optical undulation – and the experimental data he had gathered on diffraction of white light. Only then, in my view, did he seek a way to derive this value by recombining data skilfully selected from the *Opticks*, in order to make his demonstration more convincing, suggesting that the model had been constructed independently of the experiment. Note that Fresnel immediately named this concept "longueur d'ondulation", by the way (Fresnel 1866, I: 18)[6], and straightforwardly inserted it in a context where it was a key feature to put his own first version of what would become the principle of interference of light to the test.

Be that as it may, the version of his "mémoire on diffraction" that was finally published in March 1816 (known as the "second mémoire") includes a new series of experiments which were carried out with the help of François Arago in the meantime, using monochromatic light identified as "intermediate between orange and red", to which Fresnel now assigns a "longueur d'ondulation" of 623 nm, "deduced [he says] from Newton's observations of the coloured rings" (Fresnel 1866, I: 101; 107); and thus identical to that proposed by Young (see Figure 1).

Shortly afterwards, Fresnel repeatedly, and without justification, associated a "longueur d'ondulation" of 577 nm to the yellow component of the spectrum (Fresnel 1866, I: 126; 153; 168) – thus also identical to that proposed in Figure 1, but still without connecting this value to Young's work. And it was not until 1818 that Fresnel assigned a completely new value of 638 nm to the spatial period of the red spectral component, which can finally be considered his own (Fresnel 1866, I: 326).

Having now noted, on the one hand, the difference between the values of lengths given by Young and those that could actually be obtained following Newton's data to the letter and, on the other hand, the difficulty in determining the reasons for correcting Newton's data and arrive precisely at Young's, I feel justified in concluding that, from 1816 (when Fresnel probably read Young with Arago's help) until 1818, it was the values of the lengths of undulation established by Young that Fresnel employed without attributing them to their author. Consequently, despite the limited success of Young's optical theory at the time of its publication as compared to that later encountered by Fresnel's, despite the growing distance between those two theories as Fresnel kept on improving his (Buchwald 1989), and despite the huge disparities which shall be noted between Young's longitudinal, mechanical undulation of ether of an indeterminate form, and the transverse, sinusoidal electromagnetic wave of modern optics, the *length of an undulation of light as Thomas Young invented it* eventually had an impact on the subsequent development of the concept of optical wavelength; at least as his first determination of the values of the lengths of the undulations bearing the different sensations of colours served as a stake to Fresnel's explorations.

## Competing Interests.

The author declares that he has no financial or non-financial personal interests directly or indirectly related to this work.

**Funding**

The research for this paper did not receive any specific funding, but was supported by a portion of the recurring funds allocated annually to the author's laboratory by Aix-Marseille Université and the CNRS.

**Data Accessibility**

This article contains no additional data.

---

[6] Later, Fresnel would actually write: « C'est ce qu'on entend ordinairement par largeur [breadth] de l'onde, quand on parle des ondes qui se forment à la surface d'un liquide. Mais j'appelle ici longueur [length] de l'onde ou longueur d'ondulation l'intervalle compris entre le premier et le dernier point ébranlé dans le fluide par une oscillation de la particule vibrante » (Fresnel 1866, II: 45).

**Declaration of AI Use**
No generative AI tool was used in the course of writing this paper.